\documentclass[paper]{JHEP3}

\usepackage{latexsym}
\usepackage{amsmath}
\usepackage{amscd}
\usepackage{amsfonts}
\usepackage{amsthm}
\usepackage{graphicx}
\allowdisplaybreaks[1]

\def\IN{\mathbb{N}}
\def\IZ{\mathbb{Z}}
\def\IR{\mathbb{R}}
\def\slash{\llap /}
\def\slashed#1{#1\!\slash\,}
\newcommand{\D}{\mathcal{D}}    
\newcommand{\Tr}{{\rm Tr}}      
\def\ap#1{\alpha^{\prime\,#1}}
\def\Dpartial{{\cal D}}

\title{Testing the fermionic terms in the non-abelian D-brane effective action
through order $\alpha'{}^3$}
\author{Mees de Roo and Martijn G.C. Eenink\\
Institute for Theoretical Physics\\
   Nijenborgh 4, 9747 AG Groningen,
     The Netherlands\\
     E-mail: \email{m.de.roo@phys.rug.nl}, \email{m.g.c.eenink@phys.rug.nl}}
\author{Paul Koerber\thanks{Aspirant FWO} and Alexander
Sevrin\\
    Theoretische Natuurkunde, Vrije Universiteit Brussel \\
    Pleinlaan 2, B-1050 Brussels, Belgium\\
        E-mail: \email{koerber@tena4.vub.ac.be}, \email{asevrin@tena4.vub.ac.be}}

\preprint{UG-02/40, VUB/TENA/02/04\\ \hepth{0207015}}

\abstract{Recently the construction of the non-abelian effective
D-brane action was performed through order $\alpha'{}^3$ including the terms
quadratic in the gauginos. This result can be tested by calculating the spectrum in
the presence of constant magnetic background fields and comparing it to the
string theoretic predictions. This test was already performed for the
purely bosonic terms. In this note we extend the test to the fermionic
terms. We obtain perfect agreement.}


\begin{document}
\section{Introduction}

While the abelian tree-level effective action for D$p$-branes is known through all orders in
$\alpha'$, at least in the limit of slowly varying fields, this is not so in the non-abelian case.
In leading order the effective action for $n$ coinciding D$p$-branes is the ten-dimensional $N=1$ supersymmetric
$U(n)$ Yang-Mills theory dimensionally reduced to $p+1$ dimensions. There are no ${\cal O}(\alpha')$ corrections.
The bosonic ${\cal O}(\alpha'{}^2)$ were first obtained in \cite{direct} and \cite{direct1}
while the fermionic terms were obtained
in \cite{goteborg} and \cite{bilal}.
In \cite{goteborg} supersymmetry fixed the correction while in \cite{bilal} a direct calculation starting from
four-point open superstring amplitudes was used. Requiring the existence of certain BPS configurations
allowed for the determination of the bosonic ${\cal O}(\alpha'{}^3)$ terms in the effective action \cite{sk1}.
Just recently, in \cite{groningen}, supersymmetry was used not only to confirm the results of \cite{sk1} but to
construct the terms quadratic in the gauginos through this order as well.

Lacking direct string theoretic calculations, checks of these results are called for.
In \cite{HT}, further developed in \cite{DST} and \cite{STT}, such a test was proposed.
One starts from two D$2p$-branes wrapped around a $p$-dimensional torus. When switching on constant
magnetic background fields this yields, upon T-dualizing, two intersecting D$p$-branes. String theory 
allows for the calculation of the spectrum of strings stretching between different branes \cite{callan}, \cite{leigh}. In
the context of the effective action, the spectrum should be reproduced by the mass spectrum of the
off-diagonal field fluctuations. In \cite{test} it was shown that the bosonic terms through ${\cal O}(\alpha'{}^3)$
correctly reproduce the spectrum of the gauge fields. In the present paper we will extend this analysis to
the terms quadratic in the gauginos. Throughout the paper we will put $2\pi\alpha'=1$ and we will follow the conventions of \cite{groningen}.
\section{The spectrum from string theory}
\label{string}
We consider a constant magnetic background on two coincident D$2p$-branes,
\begin{eqnarray}
{\cal F}_{2a-1\,2a}=i
\left(
   \begin{array}{cc}
     f_a & 0 \\
     0 & -f_a
   \end{array}
  \right),
\end{eqnarray}
with $a\in\{1,2,\cdots,p\}$ and $f_a\in\IR$, $f_a>0$. We choose a gauge
such that $ {\cal A}_{2a-1}=0$, $\forall a$, and T-dualize in the $2$, $4$, ..., $2p$ directions.
We end up with two intersecting D$p$-branes. We want to calculate the spectrum of open strings stretching between
the two branes. We take the first brane
located along the 1, 3, ..., $2p-1$ directions. The other brane has been
rotated with respect to the first one over an angle $\theta_1$ in the $12$ plane, over an angle $\theta_2$ in the $34$ plane, ...,
over an angle $\theta_p$ in the $2p-1\,2p$ plane. The angles are determined by the magnetic fields,
\begin{eqnarray}
\theta_a=2\arctan f_a,\quad \forall a \in\{1,2,\cdots, p\}.
\end{eqnarray}   
Inspired by \cite{leigh}, we introduce,
\begin{eqnarray}
&&\hat X{}^{2a-1}=\cos \theta_a X^{2a-1}+\sin \theta_a X^{2a},\quad
\hat X{}^{2a}=-\sin \theta_a X^{2a-1}+\cos \theta_a X^{2a}, \nonumber\\
&& \hat \psi{}_\pm^{2a-1}=\cos \theta_a \psi_\pm^{2a-1}+\sin \theta_a \psi_\pm^{2a},\quad
\hat \psi{}_\pm^{2a}=-\sin \theta_a \psi_\pm^{2a-1}+\cos \theta_a \psi_\pm^{2a},
\end{eqnarray}
we impose the boundary conditions,
\begin{eqnarray}
\mbox{at }\sigma=0:&& \partial_\sigma X^{2a-1}=0,\quad \partial_\tau X^{2a}=0, \nonumber\\
&& \psi_+^{2a-1}=\psi_-^{2a-1},\quad \psi_+^{2a}=-\psi_-^{2a}; \nonumber\\
\mbox{at }\sigma=\pi:&& \partial_\sigma \hat X{}^{2a-1}=0,\quad \partial_\tau \hat X{}^{2a}=0, \nonumber\\
&& \hat\psi{}_+^{2a-1}=\eta\hat\psi{}_-^{2a-1},\quad \hat\psi{}_+^{2a}=-\eta\hat\psi{}_-^{2a},
\end{eqnarray}
where $\eta=+1$ or $\eta=-1$ in the Ramond and the Neveu-Schwarz sector resp.
Upon solving the equations of motion and implementing the boundary conditions we get the following
expansion for the bosons,
\begin{eqnarray}
X^{2a-1}&=& \frac{i}{\sqrt{2\pi}}\sum_{n\in\IZ} \left(
\frac{\alpha_{n_{+a}}}{n_{+a}}e^{-in_{+a}\tau}\cos{n_{+a}\sigma}+
\frac{\alpha_{n_{-a}}}{n_{-a}}e^{-in_{-a}\tau}\cos{n_{-a}\sigma}
\right), \nonumber\\
X^{2a}&=& \frac{i}{\sqrt{2\pi}}\sum_{n\in\IZ} \left(
\frac{\alpha_{n_{+a}}}{n_{+a}}e^{-in_{+a}\tau}\sin{n_{+a}\sigma}-
\frac{\alpha_{n_{-a}}}{n_{-a}}e^{-in_{-a}\tau}\sin{n_{-a}\sigma}
\right) ,
\end{eqnarray}
where we introduced
\begin{eqnarray}
\varepsilon_a\equiv \frac{\theta_a}{\pi},\qquad n_{\pm a}\equiv n\pm \varepsilon_a\mbox{ with }n\in\IZ.
\end{eqnarray}
In the Ramond sector (we do not need the Neveu-Schwarz sector for this paper), we get
\begin{eqnarray}
\psi_\pm^{2a-1}&=&\frac 1 2 \sum_{n\in\IZ}\left(
d_{n_{+a}}e^{-in_{+a}(\tau\pm\sigma)}+d_{n_{-a}}e^{-in_{-a}(\tau\pm\sigma)}\right), \nonumber\\
\psi_\pm^{2a}&=&\pm \frac i 2 \sum_{n\in\IZ}\left(
d_{n_{+a}}e^{-in_{+a}(\tau\pm\sigma)}-d_{n_{-a}}e^{-in_{-a}(\tau\pm\sigma)}
\right).
\end{eqnarray}
The non-vanishing (anti-)commutation relations are
\begin{eqnarray}
&&[\alpha_{m_{+a}},\alpha_{n_{-b}}]=m_{+a}\delta_{m+n}\delta_{ab}, \nonumber\\
&& \{d_{m_{+a}},d_{n_{-b}} \}=\delta_{m+n}\delta_{ab}.
\end{eqnarray}
Both $X^{2a-1}$ and $X^{2a}$ contribute to the vacuum energy (in units where $2\pi\alpha'=1$) by
$-\pi/12+\pi\varepsilon_a(1-\varepsilon_a)/2$ which is precisely cancelled by the contribution of the Ramond
fermions. So just as for the case without magnetic fields, the vacuum energy vanishes in the Ramond sector.
The (light-cone) states which in the absence of magnetic fields reduce to the gauginos are of the form
\begin{eqnarray}
\label{staat}
\prod_{a=1}^p(\alpha_{-\varepsilon_a})^{m_a}(d_{-\varepsilon_a})^{l_a}|0\rangle ,
\end{eqnarray} 
where $m_a\in\IN$ and $l_a\in\{0,1\}$, $\forall a\in\{1,\cdots,p\}$
and $|0\rangle$ carries a chiral spinor representation of $Spin(8-2p)$. 
Their masses are given by
\begin{eqnarray}
\label{massa}
M^2=\sum_{a=1}^p2\left(
m_a+l_a
\right)\theta_a.
\end{eqnarray}
\section{The spectrum from the effective action}
\label{leading}
\subsection{The leading term}
To set the stage we will first review some of the results of \cite{spec1} and \cite{spec2}.
Our starting point is the $U(2)$ $d=10$ $N=1$ supersymmetric
Yang-Mills theory\footnote{The calculation of the spectrum only probes
$U(2)$ sub-sectors of the full $U(n)$ theory \cite{spec2}. Note that we always write
spacetime indices as lower indices.},
\begin{eqnarray}
\label{SYMLagr}
{\cal   L}_{0} \;=\; -\frac{1}{g^2}
  \Tr\, \{ -\tfrac{1}{4}F_{ab}F_{ab}+\tfrac{1}{2}\bar{\chi}\slashed{\D}\chi \}
\,.
\end{eqnarray}
For simplicity we will put $g=1$ throughout this paper. Compactifying $2p$ dimensions on a torus,
we introduce complex coordinates for the compact directions, $z^\alpha=(x^{2\alpha-1}+i x^{2\alpha})/\sqrt{2}$,
$\bar{z}^{\bar\alpha}=(z^\alpha)^*$, $\alpha\in\{1,\cdots,p\}$. We switch
on constant magnetic background fields in the compact directions $ {\cal F}_{\alpha\beta}=
{\cal F}_{\bar\alpha\bar\beta}=0$, ${\cal F}_{\alpha\bar\beta}=0$ for $\alpha\neq\beta$ and\footnote{We do not
sum over repeated indices corresponding to {\em complex} coordinates, unless indicated otherwise.}
\begin{eqnarray}
{\cal F}_{\alpha\bar\alpha}=
i\left(
   \begin{array}{cc}
     f_\alpha & 0 \\
     0 & -f_\alpha
   \end{array}
  \right),
\end{eqnarray}
where the $f_\alpha$, $\alpha\in\{1,\cdots,p\}$ are imaginary
constants such that $if_\alpha>0$.
We only consider the off-diagonal components of the fermions,
\begin{eqnarray}
\chi=
i\left(
   \begin{array}{cc}
     0 & \chi^+ \\
     \chi^- & 0
   \end{array}
  \right),
\end{eqnarray}
as the diagonal fluctuations probe the abelian part of the action.
Using the previous choices, we can rewrite the second term in eq.\ (\ref{SYMLagr}) as,
\begin{eqnarray}
\label{LeOr}
{\cal L}_{\text{\em fermion}}=\bar\chi^- \left(\partial\!\!\!/_{NC}+ {\cal D}\!\!\!\!/\right)\chi^+,
\end{eqnarray}
where subindex $NC$ denotes operators acting in the non-compact directions only and $ {\cal D }\equiv \partial+2i {\cal A}$,
with $ {\cal A}$ the background gauge fields. The covariant derivatives satisfy
\begin{eqnarray}
\label{comm}
[{\cal D}_\alpha,{\cal D}_{\bar\beta}]=2i\delta_{\alpha\beta}f_\alpha.
\end{eqnarray}
The equations of motion readily follow from eq.\ (\ref{LeOr}),
\begin{eqnarray}
\label{LeOreom}
\left(\partial\!\!\!/_{NC}+ {\cal D}\!\!\!\!/\right)\chi^+=0.
\end{eqnarray}
Squaring the kinetic operator in eq.\ (\ref{LeOreom}) and using eq. (\ref{comm}), we get,
\begin{eqnarray}
\label{eomsq}
\left(\Box_{NC}+2\sum_{\alpha=1}^{p}\left\{
{\cal D}_\alpha {\cal D}_{\bar\alpha}-if_\alpha-if_\alpha\gamma_{\alpha\bar\alpha}
\right\}\right)\chi^+=0,
\end{eqnarray}
where
$\gamma_{\alpha\bar\alpha}\equiv(\gamma_{\alpha}\gamma_{\bar\alpha}-\gamma_{\bar\alpha}\gamma_{\alpha})/2$,
$(\gamma_{\alpha\bar\alpha})^2=1$. Once a
complete set of eigenfunctions is constructed for the second part
in eq.\ (\ref{eomsq}), we can bring the relation above in the form
$(\Box-M^2)\chi=0$ and read off the mass $M$. Such eigenfunctions
are obtained from a spinor $|0 \rangle$ satisfying $ {\cal
D}_{\bar\alpha}|0 \rangle=0$, $\forall \alpha$, which has been explicitly
constructed in \cite{spec1} and \cite{spec2}. We now introduce the
complete set of functions
$|\{(m_1,n_1),\,(m_2,n_2),\,\cdots(m_p,n_p)\} \rangle$,
$m_1,\,m_2,\,\cdots , m_p\in\IN$ and $n_1,\,n_2,\,\cdots
n_p\in\{-1,+1\}$ by
\begin{eqnarray}
&&|\{(m_1,n_1),\,(m_2,n_2),\,\cdots , (m_p,n_p)\}\rangle\equiv \nonumber\\
&&\quad\frac 1 2 (1+n_1\gamma_{1\bar 1})\frac 1 2 (1+n_2\gamma_{2\bar 2})\cdots\frac 1 2 (1+n_p\gamma_{p\bar p})
{\cal D}_1^{m_1}{\cal D}_2^{m_2}\cdots{\cal D}_p^{m_p}|0\rangle.
\end{eqnarray}
Expanding the fermion,
\begin{eqnarray}
\chi^+(y,z,\bar z)=\sum_{\{(m,n)\}}\chi^+_{\{(m,n)\}}(y)|\{(m,n)\}\rangle,
\end{eqnarray}
where $\{(m,n)\}\equiv\{(m_1,n_1),\,(m_2,n_2),\,\cdots(m_p,n_p)\}$ and
$y$ collectively denotes the non-compact coordinates.
Using this, one gets from eq.\ (\ref{comm}) and eq.\ (\ref{eomsq}) that the mass  of
$\chi^+_{\{(m,n)\}}(y)$
is given by
\begin{eqnarray}
\label{masslo}
M^2=2i\sum_{\alpha=1}^p\left(
2m_\alpha+1+n_\alpha
\right)f_\alpha.
\end{eqnarray}
Replacing $f_\alpha$ by $\mbox{arctanh}(f_\alpha)$ in eq.\ (\ref{masslo}) yields the stringy result, eq. (\ref{massa}).
As expected, we only get agreement for very small
magnetic background fields. The higher order terms in the effective action
should add to this such that the string result gets reproduced. In particular one notices from this that only even orders
in $\alpha'$ contribute to the spectrum.
\subsection{The ${\cal O}(\alpha'{}^2)$ contribution to the spectrum}
Modulo field redefinitions and up to terms terms quartic in the fermions, 
the effective action through ${\cal O}(\alpha'{}^2)$ is
given by $ {\cal L}={\cal L}_0+{\cal L}_2$ where ${\cal L}_0$ was given in eq.\ (\ref{SYMLagr}) and ${\cal L}_2$
is given by \cite{direct}, \cite{direct1}, \cite{goteborg},
\begin{eqnarray}
\label{Lag2}
{\cal L}_2&=&\mbox{STr} \bigg(
x_1F_{ab}F_{ab}F_{cd}F_{cd}+x_2F_{ab}F_{bc}F_{cd}F_{da} \nonumber\\
&&+x_3 F_{ab}F_{ac}\bar\chi\gamma_b {\cal D}_c\chi
+x_4 F_{ab} {\cal D}_{a}F_{cd}\bar\chi\gamma_{bcd} \chi
\bigg),
\end{eqnarray}
where STr denotes the symmetrized trace and
\begin{eqnarray}
x_1=- \frac{1}{32},\qquad x_2 =\frac 1 8, \qquad x_3 = -\frac 1 4,\qquad x_4=- \frac{1}{16}.
\end{eqnarray}
Again we want to calculate the fermionic spectrum through this order. It is clear that, as the background
magnetic fields are (covariantly) constant, only the term proportional to $x_3$ will contribute.
Following exactly the same strategy as above, we get the equations of motion,
\begin{eqnarray}
\left(\partial\!\!\!/_{NC}+ {\cal D}\!\!\!\!/-\frac{2 x_3}{3}\sum_{\alpha=1}^p
f_\alpha^2\left(\gamma_\alpha {\cal D}_{\bar\alpha}+\gamma_{\bar\alpha} {\cal D}_\alpha\right)\right)\chi^+=0.
\end{eqnarray}
Again squaring the kinetic operator we get,
\begin{eqnarray}
\left(\Box_{NC}+2\sum_{\alpha=1}^p(1- \frac{4x_3}{3}f_\alpha^2)\left\{
{\cal D}_\alpha {\cal D}_{\bar\alpha}-if_\alpha-if_\alpha\gamma_{\alpha\bar\alpha}
\right\}\right)\chi^+=0,
\end{eqnarray}
where we ignored terms proportional to $f^4$ as they are of higher order in $\alpha'$. However such terms
will be relevant for a test of the, as of yet still unknown, ${\cal O}(\alpha'{}^4)$ terms in the
effective action. It is clear that this gives the same
spectrum as in eq.\ (\ref{masslo}), but with $f_\alpha$ replaced by,
\begin{eqnarray}
f_\alpha\rightarrow f_\alpha - \frac{4 x_3}{3}f_\alpha^2.
\end{eqnarray}
Consistency with the string spectrum requires that $x_3=-1/4$ which agrees with the result
based on supersymmetry arguments and the direct calculation from open superstring amplitudes \cite{goteborg}, \cite{bilal}.

In \cite{STT} it was shown that demanding that the spectrum of the gauge fields is correctly reproduced, combined with
the requirement that the abelian limit agrees with the known result, completely fixes the bosonic part
of the effective action through order $\alpha'{}^2$.
It is clear from the above that this is not the case for the fermionic terms which already indicates that the spectral
test is indeed weaker for the terms containing fermions than for the purely bosonic terms.
\subsection{Testing the ${\cal O}(\alpha'{}^3)$ terms}
At order ${\cal O}(\alpha'{}^3)$ the effective action is given by
$ {\cal L}={\cal L}_0+{\cal L}_2+{\cal L}_3$, where $ {\cal L}_0$
and $ {\cal L}_2$ are given in eq.\ (\ref{SYMLagr}) and eq.\
(\ref{Lag2}). The last term is given by \cite{groningen}\footnote{
We took $U(n)$ generators in the fundamental representation satisfying 
$[t^X,t^Y]=f^{XYZ}t^Z$ where $f^{XYZ}$ is completely anti-symmetric
and $\Tr (t^Xt^Y)=-\delta_{XY}$.},
\cite{sk1},
\begin{eqnarray}
{\cal L}_3 &=&- \frac{\zeta(3)}{16\pi^3} f^{XYZ}f^{VWZ}\bigg[2\,F_{ab}{}^{X}F_{cd}{}^{W}
\Dpartial_eF_{bc}{}^{V}\Dpartial_eF_{ad}{}^{Y}
-2\,F_{ab}{}^{X}F_{ac}{}^{W}
\Dpartial_dF_{be}{}^{V}\Dpartial_dF_{ce}{}^{Y}
\nonumber\\
&&\qquad +F_{ab}{}^{X}F_{cd}{}^{W}
\Dpartial_eF_{ab}{}^{V}\Dpartial_eF_{cd}{}^{Y}
\nonumber\\
&&\qquad -4\,F_{ab}{}^{W}
\Dpartial_cF_{bd}{}^{Y}\bar\chi^{X}\gamma_{a}\,\Dpartial_d\Dpartial_c\chi^{V}
-4\,F_{ab}{}^{W}
\Dpartial_cF_{bd}{}^{Y}\bar\chi^{X}\gamma_{d}\,\Dpartial_a\Dpartial_c\chi^{V}
\nonumber\\
&&\qquad +2\,F_{ab}{}^{W}
\Dpartial_cF_{de}{}^{Y}\bar\chi^{X}\gamma_{ade}\,\Dpartial_b\Dpartial_c\chi^{V}
+2\,F_{ab}{}^{W}
\Dpartial_cF_{de}{}^{Y}\bar\chi^{X}\gamma_{abd}\,\Dpartial_e\Dpartial_c\chi^{V}
\bigg]
\nonumber\\
&& - \frac{\zeta(3)}{16\pi^3}\ f^{XYZ}f^{UVW}f^{TUX}\bigg[4\,
  F_{ab}{}^{Y}F_{cd}{}^{Z}F_{ac}{}^{V}F_{be}{}^{W}F_{de}{}^{T}
+2\,F_{ab}{}^{Y}F_{cd}{}^{Z}F_{ab}{}^{V}F_{ce}{}^{W}F_{de}{}^{T}
\nonumber\\
&&\qquad -11\,F_{ab}{}^{Y}F_{cd}{}^{Z}F_{cd}{}^{V}\bar\chi^{T}\gamma_{a}\,
\Dpartial_b\chi^{W}
+22\,F_{ab}{}^{Y}F_{cd}{}^{Z}F_{ac}{}^{V}\bar\chi^{T}\gamma_{b}\,
\Dpartial_d\chi^{W}
\nonumber\\
&&\qquad +18\,F_{ab}{}^{Y}F_{cd}{}^{V}F_{ac}{}^{W}\bar\chi^{T}\gamma_{b}\,
\Dpartial_d\chi^{Z}
+12\,F_{ab}{}^{T}F_{cd}{}^{Y}F_{ac}{}^{V}\bar\chi^{Z}\gamma_{b}\,
\Dpartial_d\chi^{W}
\nonumber\\
&&\qquad +28\,F_{ab}{}^{T}F_{cd}{}^{Y}F_{ac}{}^{V}\bar\chi^{W}\gamma_{b}\,
\Dpartial_d\chi^{Z}
-24\,F_{ab}{}^{Y}F_{cd}{}^{V}F_{ac}{}^{T}\bar\chi^{W}\gamma_{b}\,
\Dpartial_d\chi^{Z}
\nonumber\\
&&\label{L3}\qquad +8\,F_{ab}{}^{T}F_{cd}{}^{Y}F_{ac}{}^{Z}\bar\chi^{V}\gamma_
{b}\,
\Dpartial_d\chi^{W}
-12\,F_{ab}{}^{T}F_{ac}{}^{Y}
\Dpartial_bF_{cd}{}^{V}\bar\chi^{Z}\gamma_{d}\,\bar\chi^{W}
\\
&&\qquad -8\,F_{ab}{}^{Y}F_{ac}{}^{T}
\Dpartial_bF_{cd}{}^{V}\bar\chi^{Z}\gamma_{d}\,\bar\chi^{W}
+22\,F_{ab}{}^{V}F_{ac}{}^{Y}
\Dpartial_bF_{cd}{}^{T}\bar\chi^{Z}\gamma_{d}\,\bar\chi^{W}
\nonumber\\
&&\qquad -4\,F_{ab}{}^{Y}F_{cd}{}^{T}
\Dpartial_eF_{ac}{}^{V}\bar\chi^{Z}\gamma_{bde}\,\bar\chi^{W}
+4\,F_{ab}{}^{Y}F_{ac}{}^{T}
\Dpartial_cF_{de}{}^{V}\bar\chi^{Z}\gamma_{bde}\,\bar\chi^{W}
\nonumber\\
&&\qquad +4\,F_{ab}{}^{T}F_{cd}{}^{Y}F_{ce}{}^{V}\bar\chi^{Z}\gamma_{abd}\,
\Dpartial_e\chi^{W}
-8\,F_{ab}{}^{Y}F_{cd}{}^{T}F_{ce}{}^{V}\bar\chi^{Z}\gamma_{abd}\,
\Dpartial_e\chi^{W}
\nonumber\\
&&\qquad +6\,F_{ab}{}^{V}F_{cd}{}^{Y}F_{ce}{}^{W}\bar\chi^{Z}\gamma_{abd}\,
\Dpartial_e\chi^{T}
+5\,F_{ab}{}^{V}F_{cd}{}^{W}F_{ce}{}^{Y}\bar\chi^{Z}\gamma_{abd}\,
\Dpartial_e\chi^{T}
\nonumber\\
&&\qquad +6\,F_{ab}{}^{Y}F_{ac}{}^{T}F_{de}{}^{V}\bar\chi^{Z}\gamma_{bcd}\,
\Dpartial_e\chi^{W}
-2\,F_{ab}{}^{Y}F_{ac}{}^{T}F_{de}{}^{Z}\bar\chi^{V}\gamma_{bcd}\,
\Dpartial_e\chi^{W}
\nonumber\\
&&\qquad +4\,F_{ab}{}^{Y}F_{ac}{}^{V}F_{de}{}^{Z}\bar\chi^{W}\gamma_{bcd}\,
\Dpartial_e\chi^{T}
+4\,F_{ab}{}^{T}F_{cd}{}^{V}F_{ce}{}^{Y}\bar\chi^{Z}\gamma_{abd}\,
\Dpartial_e\chi^{W}
\nonumber\\
&&\qquad -4\,F_{ab}{}^{Y}F_{cd}{}^{V}F_{ce}{}^{W}\bar\chi^{Z}\gamma_{abd}\,
\Dpartial_e\chi^{T}
\nonumber\\
&&\qquad +\tfrac{1}{2}\,F_{ab}{}^{Y}F_{cd}{}^{T}F_{ef}{}^{V}\bar\chi^{Z}\gamma_
{abcde}\,
\Dpartial_f\chi^{W}
+\tfrac{1}{2}\,F_{ab}{}^{Y}F_{cd}{}^{T}
   F_{ef}{}^{Z}\bar\chi^{V}\gamma_{abcde}\,
\Dpartial_f\chi^{W}\bigg]\,.
\nonumber
\end{eqnarray}
The overall multiplicative constant can not be fixed by the methods used
in either \cite{sk1} or
\cite{groningen}. It gets determined through comparison with the relevant higher order
derivative terms obtained in \cite{bilal1}.

We now turn to the calculation of the spectrum. It is clear that terms involving a derivative
on the field-strength will not contribute as the field-strength is covariantly constant. Furthermore,
any term having two field-strengths contracted with a single $f$-symbol can be ignored as well as we
took the background field-strength in the Cartan subalgebra of $SU(2)$. Having discarded these terms
we note that for the remaining terms the group theoretical factors, for our particular choice of background, are such
that the Lie algebra indices on the gauginos are anti-symmetric when interchanging them. This implies
that all terms involving a single or five gamma-matrices will vanish 
(up to a total derivative) as well. The only terms which can potentially contribute
are now proportional to
\begin{eqnarray}
\label{leftover}
&&\qquad x_1\,F_{ab}{}^{T}F_{cd}{}^{Y}F_{ce}{}^{V}\bar\chi^{Z}\gamma_{abd}\,
\Dpartial_e\chi^{W}
+x_2\,F_{ab}{}^{Y}F_{cd}{}^{T}F_{ce}{}^{V}\bar\chi^{Z}\gamma_{abd}\,
\Dpartial_e\chi^{W}
\nonumber\\
&&\qquad +x_3\,F_{ab}{}^{Y}F_{ac}{}^{T}F_{de}{}^{V}\bar\chi^{Z}\gamma_{bcd}\,
\Dpartial_e\chi^{W}
+x_4\,F_{ab}{}^{T}F_{cd}{}^{V}F_{ce}{}^{Y}\bar\chi^{Z}\gamma_{abd}\,
\Dpartial_e\chi^{W},
\end{eqnarray}
where
\begin{eqnarray}
\label{resgr}
x_1=+4,\qquad x_2=-8,\qquad x_3=+6,\qquad x_4=+4.
\end{eqnarray}
Rewriting eq.\ (\ref{leftover}) in terms of the background and the off-diagonal
fermions we get a result proportional to
\begin{eqnarray}
(x_1+x_2+x_4)\sum_{\alpha=1}^p\sum_{\beta=1}^p
f_\beta f_\alpha^2\left(
\bar\chi^-\gamma_{\bar\beta\beta\alpha} {\cal D}_{\bar\alpha}\chi^+ +
\bar\chi^-\gamma_{\bar\beta\beta\bar\alpha} {\cal D}_{\alpha}\chi^+
\right),
\end{eqnarray}
which indeed vanishes when using eq.\ (\ref{resgr}).
\section{Conclusions}
Though the spectral test is not as restrictive for the fermionic terms as it was
for the purely bosonic terms, it is still gratifying to see that the fermionic terms pass
it as well. The present proposal for the effective action through ${\cal O}(\alpha'{}^3)$
and up to terms quartic in the gauginos, is of the form,
\begin{eqnarray}
{\cal L}={\cal L}_0+{\cal L}_2+{\cal L}_3+ {\cal O}(\alpha'{}^4),
\end{eqnarray}
where $ {\cal L}_0$, $ {\cal L}_2$ and
$ {\cal L}_3$ are given in eqs.\ (\ref{SYMLagr}), (\ref{Lag2}) and (\ref{L3}). 
The purely bosonic part of $ {\cal L}_3$, which was obtained in \cite{sk1}, passed the spectral test in \cite{test} while
other proposals in the literature for these terms failed to do so. A very strong test of
the purely bosonic terms was provided by the results in \cite{groningen} where the supersymmetry invariant
at order $\alpha'{}^3$ was constructed. Its bosonic part precisely matches the one obtained in \cite{sk1}.
In addition, the terms quadratic in the gauginos were obtained as well. 
The bosonic and fermionic terms were tested in \cite{groningen} by checking the closure of the commutator of
two supersymmetry transformations. Furthermore, as was shown in the present paper,
the fermionic terms correctly reproduce the gaugino spectrum in the presence of magnetic backgrounds.

So at this point there is no doubt left that we do have the correct description of the non-abelian D-brane
effective action through order $\alpha'{}^3$ and up to and including terms quadratic in the gauginos.

\bigskip

\acknowledgments

\bigskip

The work of MGCE is part of the research programme of the ``Stichting 
voor Fundamenteel Onderzoek van de Materie'' (FOM).
PK and AS are supported in part by the ``FWO-Vlaanderen'' through project
G.0034.02 and in part by the Federal Office for Scientific, Technical 
and Cultural Affairs through the Interuniversity Attraction Pole P5/27. 
The authors were supported in part by the European
Commission RTN programme HPRN-CT-2000-00131, in which MdR and MGCE
are associated to the University of Utrecht and PK and AS are
associated to the University of Leuven. PK and AS thank the
University of Groningen for hospitality.

\end{document}